\documentclass[twocolumn,showpacs,preprintnumbers,amsmath,amssymb]{revtex4-1}

\usepackage{graphicx}
\usepackage{dcolumn}
\usepackage{bm}
\usepackage{color}

\begin{document}

\preprint{APS/123-QED}

\title{Superlattice formation lifting degeneracy protected by non-symmorphic symmetry through a metal-insulator transition in RuAs}

\author{Hisashi Kotegawa$^{1}$, Keiki Takeda$^{2}$, Yoshiki Kuwata$^{1}$, Junichi Hayashi$^{2}$, Hideki Tou$^{1}$, Hitoshi Sugawara$^{1}$, Takahiro Sakurai$^{3}$, Hitoshi Ohta$^{1,4}$, and Hisatomo Harima$^{1}$}

\affiliation{
$^{1}$Department of Physics, Kobe University, Kobe 658-8530, Japan \\
$^{2}$Muroran Institute of Technology, Muroran, Hokkaido 050-8585, Japan \\
$^{3}$Research Facility Center for Science and Technology, Kobe University, Kobe, Hyogo 657-8501, Japan\\
$^{4}$Molecular Photoscience Research Center, Kobe University, Kobe, Hyogo 657-8501, Japan
}

\date{\today}

\begin{abstract}

The single crystal of RuAs obtained by Bi-flux method shows obvious successive metal-insulator transitions at $T_{\rm MI 1}\sim255$ K and $T_{\rm MI 2}\sim195$ K. The X-ray diffraction measurement reveals a formation of superlattice of $3\times 3\times 3$ of the original unit cell below $T_{\rm MI 2}$, accompanied by a change of the crystal system from the orthorhombic structure to the monoclinic one. Simple dimerization of the Ru ions is nor seen in the ground state. The multiple As sites observed in nuclear quadrupole resonance (NQR) spectrum also demonstrate the formation of the superlattice in the ground state, which is clarified to be nonmagnetic. The divergence in $1/T_1$ at $T_{\rm MI 1}$ shows that a symmetry lowering by the metal-insulator transition is accompanied by strong critical fluctuations of some degrees of freedom. Using the structural parameters in the insulating state, the first principle calculation reproduces successfully the reasonable size of nuclear quadrupole frequencies, $\nu_Q$ for the multiple As sites, ensuring the high validity of the structural parameters. The calculation also gives a remarkable suppression in the density of states (DOS) near the Fermi level, although the gap opening is insufficient. A coupled modulation of the calculated Ru $d$ electron numbers and the crystal structure proposes a formation of charge density wave (CDW) in RuAs. Some lacking factors remain, but it shows that a lifting of degeneracy protected by the non-symmorphic symmetry through the superlattice formation is a key ingredient for the metal-insulator transition in RuAs.

\end{abstract}

\pacs{}
\maketitle

\section{Introduction}

Metal-insulator transition derived by a drastic revolution in conductivity is an exotic phenomenon in condensed matter physics.
Except for a gap opening induced by strong electronic correlations, it is generally a cooperative symmetry breaking brought by a coupling of the Fermi surface instability and degrees of freedom in solid, such as charge, spin, orbital, or multipole.\cite{Peierls,Pytte,Khomskii,Streltsov,Harima}
Symmetry of crystal is a key ingredient to understand a picture of metal-insulator transition. 
If material is low dimensional, typically composed of one-dimensional linear chain, it generally possesses Fermi surface with a good nesting.
It promotes a Peierls transition, opening a gap at the Fermi level.\cite{Peierls} 
In case spins contribute to the Peierls transition, it is often characterized by a dimerization to produce a spin-singlet.\cite{Pytte}
If material is highly symmetric, on the other hand, the ions can be located under high local symmetry, inducing the orbital degeneracy or multipole degeneracy. 
A contribution of such degrees of freedom to a metal-insulator transition is a debate topic.\cite{Khomskii,Streltsov,Harima}


In 2012, Hirai {\it et al.} have reported that simple binary RuP and RuAs undergo a metal-insulator transition.\cite{Hirai}
They crystallize in the orthorhombic structure in the space group of $Pnma$.
The nearest-neighbor Ru ions form a zig-zag chain along the $a$ axis, while the second and third nearest-neighbor Ru ions form a zig-zag ladder along the $b$ axis. This structure is neither a simple low-dimensional one nor a high-symmetric one.
Two successive metal-insulator transitions have been reported in the polycrystalline samples,\cite{Hirai} and this behavior is more remarkable in RuP than in RuAs, and it disappears in the isostructural RuSb.
For RuAs, the transition temperatures have been estimated to be $250$ K and $190$ K, which are denoted as $T_{\rm MI1}$ and $T_{\rm MI2}$ in this paper.
Interestingly, a Rh doping to RuAs and RuP strongly suppresses the insulating state, and induces superconductivity.\cite{Hirai,Li}
Intriguing question is what triggers the metal-insulator transition of these materials.
The previous band calculation has suggested that the topology of Fermi surface is similar between RuAs and RuP, and that of RuSb differs from two compounds.\cite{Goto}
RuAs and RuP possess the degenerate flat bands near the Fermi level, which originates mainly from the $4d_{xy}$ orbitals of Ru; therefore, it is expected that the splitting of the flat band is related with the metal-insulator transition.
Charge density wave (CDW) may be realized, but the nesting property of the Fermi surface is not so clear, because they are not simple low-dimensional materials.
Photoemission measurement has not detected a typical feature of a CDW transition for the polycrystalline RuP,\cite{Sato} that is, the charge distribution of the Ru orbital is not clearly seen.
Except for the CDW scenario, the contribution of pseudo-degeneracy of the $d$ orbitals and a possibility of a spin-singlet formation have been discussed,\cite{Sato,Li} but a mechanism of the metal-insulator transition is still unsettled.

Recently, physical properties of a single crystal of RuP made by Sn-flux method have been reported.\cite{Chen,Fan} 
The crystal shows the clear successive transitions at 320 K and 270 K, but the ground state of the single crystal is metallic in contrast to the polycrystalline sample.
The structural transition originating from the formation of a superlattice has been confirmed at room temperature for a single crystal RuP,\cite{Hirai,Chen} whereas the structure of the polycrystalline sample at room temperature is still in the original $Pnma$.\cite{Rundqvist,Hirai} 
The formation of a superlattice is a key feature to reveal the origin of the metal-insulator transition of these materials, but unfortunately inconsistency of the physical properties of RuP between single and poly crystals complicates the problem.
In this paper, we focus RuAs to investigate the origin of the metal-insulator transition.
A single crystal of RuAs, whose property is similar to the polycrystalline one, was successfully obtained using Bi-flux method.

\section{Experimental procedure}

To make the single crystal, the starting materials of Ru : As : Bi = 1 : 1 : 35 were sealed in a silica tube, and they were heated up to 1050 $^\circ$C, followed by a slow cooling with a rate of $-5$ $^\circ$C/h down to 600 $^\circ$C. After a centrifugation, small single crystals with a maximum long axis of about 0.5 mm were obtained. 
The crystals made by this procedure are denoted as \#1.
Another procedure was attempted as \#2; the starting composition of Ru : As : Bi = 1 : 1 : 25, the maximum temperature of 1100 $^\circ$C, and the cooling rate of $-3$ $^\circ$C/h.
The size of crystals was similar between \#1 and \#2, but the sequence for \#2 was effective to reduce the by-product RuAs$_2$.
We also tried Sn-flux method, but it failed to yield single crystals of RuAs.
Very small single crystals of RuP are also obtained using Bi-flux method, but the resistivity does not show insulating behavior, and the overall behavior resembles that of the single crystal made by Sn-flux method.\cite{Chen}

The electrical resistivity ($\rho$) of RuAs was measured using a four-probe method, in which electrical contacts of wire were made by a spod-weld method. 
The current direction could not be confirmed owing to the smallness of the crystal, but it is expected to be perpendicular to the $a$ axis on the analogy of the Laue experiment for other crystals. 
The high pressure was applied by utilizing an indenter-type pressure cell and Daphne7474 as a pressure-transmitting medium.\cite{indenter,Murata}
Magnetic susceptibility measurement was performed by utilizing a Magnetic Property Measurement System (MPMS : Quantum Design).
X-ray diffraction measurements using a single crystal were made on a Rigaku Saturn724 diffractometer using multi-layer mirror monochromated Mo-K$\alpha$ radiation.
A small single crystal of $0.04 \times 0.04 \times 0.03$ mm was used for the measurement. The data were collected to a maximum $2\theta$ value of $\sim 62^{\circ}$ with using the angle scans.
For all structure analyses, the program suite SHELX was used for structure solution and least-squares refinement.\cite{Sheldrick}
Platon was used to check for missing symmetry elements in the structures.\cite{Spek}
We also performed nuclear quadrupole resonance (NQR) and nuclear magnetic resonance (NMR) for RuAs using $^{75}$As nucleus with a nuclear spin of $I=3/2$.
Band calculations were obtained through a full-potential LAPW (linear augmented plane wave) calculation within the LDA(local density approximation).

\section{Results and Discussion}

\subsection{Resistivity, pressure effect, and magnetization}

Figure 1(a) shows the temperature dependence of $\rho$ for a single crystal RuAs. 
The $\rho$ shows almost constant at high temperatures down to $T_{\rm MI1}=250$ K, and possesses a clear kink at $T_{\rm MI1}$, below which $\rho$ starts to increase. 
A sharp transition appears at $T_{\rm MI2}=190$ K, followed by a clear insulating behavior down to the lowest temperature.
As shown in the inset, the transition at $T_{\rm MI2}$ possesses obvious hysteresis similarly to the polycrystalline sample,\cite{Hirai} while hysteresis was not visible at $T_{\rm MI1}$. 
The increase in $\rho$ below $T_{\rm MI2}$ is much larger than that of the polycrystalline sample, indicating that a contribution from possible conductive impurities can be avoided in the measurement using the single crystal.
A fitting of the resistivity data into a typical activation form of $\rho(T) \propto \exp(E_g/2k_BT)$ between 50 K and $T_{\rm MI2}=190$ K gives the energy gap of $E_g=340$ K.
$\rho$ deviates from the simple activation form below 50 K.

\begin{figure}[htb]
\begin{center}
\includegraphics[width=0.85\linewidth]{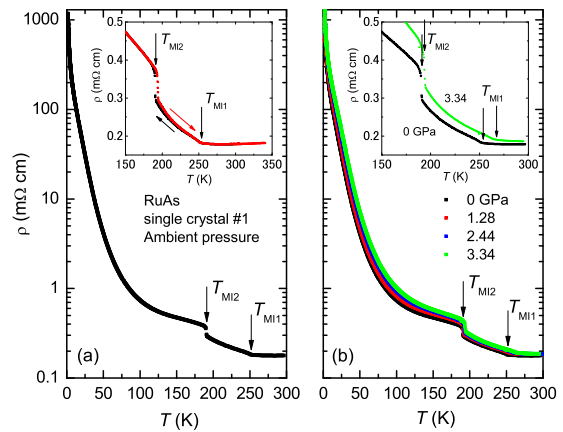}
\end{center}
\caption[]{(color online) (a) Temperature dependence of $\rho$ for RuAs at ambient pressure. The resistivity shows clearly two successive metal-insulator transitions at $T_{\rm MI1}$ and $T_{\rm MI2}$. The inset shows the hysteresis behavior at $T_{\rm MI2}$. (b) The pressure dependence of $\rho$ for RuAs. Both transition temperatures increase with elevating pressure up to 3.34 GPa.
}
\end{figure}

Figure 1(b) shows the pressure variation of $\rho$ for RuAs up to 3.34 GPa.
Both $T_{\rm MI1}$ and $T_{\rm MI2}$ increase slightly with increasing pressure.
The rates were estimated to be $+3.6$ K/GPa for $T_{\rm MI1}$ and $+0.8$ K/GPa for $T_{\rm MI2}$, respectively.
This tendency suggests that the stronger hybridization induced by a smaller volume stabilizes the insulating state, consistent with that RuP is more insulating.

\begin{figure}[htb]
\begin{center}
\includegraphics[width=0.75\linewidth]{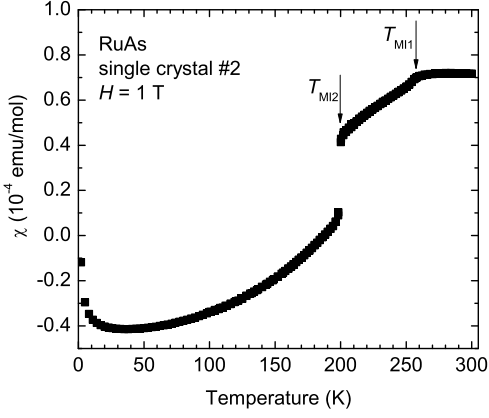}
\end{center}
\caption{Temperature dependence of the magnetic susceptibility measured for many pieces of single crystals at 1 T. Two successive transitions can be seen obviously. The $T_{\rm MI1}$ and $T_{\rm MI2}$ for \#2 are slightly higher than those of \#1.}
\label{f2}
\end{figure}

Figure 2 shows the temperature dependence of the magnetic susceptibility measured at 1 T on the cooling process.
The overall behavior is similar to that obtained in the polycrystalline sample,\cite{Hirai} but the transitions become much sharper.
The kink at $T_{\rm MI1} = 255$ K is continuous, while the discrete drop was observed at $T_{\rm MI2} = 200$ K.
The ground state is dominated by diamagnetism as well as the polycrystalline sample.\cite{Hirai} 
In the metallic state, the susceptibility is temperature independent, that is, Curie-Weiss like behavior is absent.
Here, we used the single crystal \#2, and both $T_{\rm MI1}$ and $T_{\rm MI2}$ are slightly higher than those of \#1 determined in Fig.~1.
We confirmed that there is small sample dependence among the procedures in the sample preparation.
It is speculated that they originate from the amount of deficiency of As, as described below.

\subsection{Structural analysis by X-ray diffraction measurement}

\begin{figure}[hth]
\begin{center}
\includegraphics[width=0.85\linewidth]{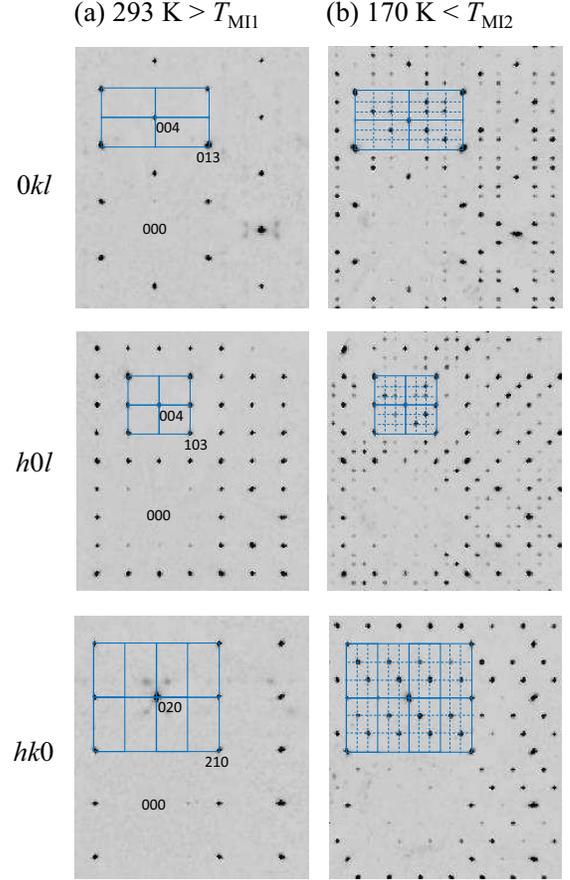}
\end{center}
\caption{(color online) Single-crystal x-ray diffraction patterns for RuAs at (a) 293 K and (b) 170 K. Reciprocal lattice vectors are shown for the orthorhombic $Pnma$ symmetry. The superlattice spots are seen below $T_{\rm MI2}$. The new periodicity can be regarded as $3 \times 3 \times 3$, but the unit cell is found to be monoclinic from the superlattice spots in the $hk0$ plane.}
\label{f2}
\end{figure}

\begin{table}[htb]
\caption{Crystallographic data of RuAs for the metallic and insulating phases.}
\begin{tabular}{lcc}\hline\hline
Temperature & 293 K & 170 K \\
Formula & RuAs & Ru$_9$As$_9$ \\
Crystal system & orthorhombic & monoclinic \\
Space group & $Pnma$ (no.62) & $P2_1/c$ (no.14) \\
$a$ (\AA) & 5.724(3) & 6.624(3) \\
$b$ (\AA) & 3.3283(14) & 18.974(7) \\
$c$ (\AA) & 6.323(3) & 8.759(4) \\
$\beta$ ($^{\circ}$) & 90 & 100.843(6) \\
$V$ (\AA$^3$) & 120.46(10) & 1082.8(8) \\
$Z$ & 4 & 4 \\
Unique reflections & 212 & 3279 \\
Residual factor $R1$ & 0.0402 & 0.0354 \\
$wR2$ & 0.0914 & 0.0785 \\
\hline\hline
\end{tabular}
\end{table}

Next, we performed a structural analysis of RuAs using a single crystal \#2.
Figure 3 shows single-crystal x-ray diffraction patterns at (a) 293 K $> T_{\rm MI1}$ and (b) 170 K $< T_{\rm MI2}$.
The crystal system above $T_{\rm MI1}$ was confirmed to be orthorhombic as reported previously.\cite{Hirai,Saparov}
In the insulating state below $T_{\rm MI2}$, the superlattice spots are clearly observed.
They appear at the positions of $0\ k/3\ l/3$ for the $0kl$ plane and $h/3\ 0\ l/3$ for the $h0l$ plane, while they are visible only at the positions of $h+1/3\ k+2/3\ 0$ or $h+2/3\ k+1/3\ 0$ for the $hk0$ plane.
As shown by the blue line, a new periodicity can be regarded as $3\times3\times3$, but the asymmetric arrangement of the superlattice spots for $hk0$ plane reveals that the unit cell below $T_{\rm MI2}$ is not orthorhombic but of monoclinic; therefore, the size of the unit cell is smaller than that of the $3\times3\times3$ cell.
In the intermediate phase between $T_{\rm MI1}$ and $T_{\rm MI2}$, on the other hand, the superlattice spots also appear (not shown), and it is likely to be incommensurate, but a detailed structural analysis has not been performed successfully yet.
Table I and II show the crystallographic data and the structural parameters for the respective sites at 293 K and 170 K obtained by the present experiments and analyses described in the section of the experimental procedure.
In the metallic state above $T_{\rm MI1}$, the space group of $Pnma$ and the lattice parameters are consistent with the previous reports for the polycrystalline samples.\cite{Hirai,Saparov}
As shown in Table II, the analysis suggests that a small amount of deficiency of As is present ($\sim1$\%).
We consider that the degree of the deficiency induces the sample dependence of the transition temperatures in RuAs, but it is not as serious as it changes the ground state drastically.

\begin{figure}[tb]
\begin{center}
\includegraphics[width=\linewidth]{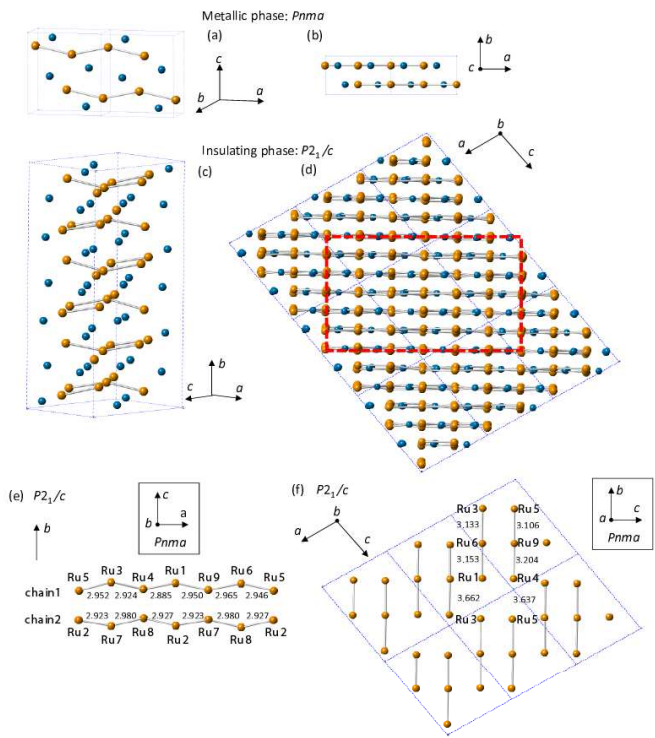}
\end{center}
\caption{(color online) The crystal structure of RuAs in the metallic phase [(a) and (b): two unit cells are shown.], and in the insulating phase [(c) and (d)]. The blue lines indicate the unit cell of each phase. The orthorhombic $Pnma$ is changed to monoclinic $P2_1/c$ at low temperatures, where new periodicity can be regarded as $3\times3\times3$ for the original unit cell. (e) Zig-zag chains of Ru ions. In $Pnma$, all the Ru ions are equivalent, and the zig-zag chain is formed by connecting the nearest-neighbor Ru ions, while two-type of chains appear in the ground state. The bond lengths of the neighboring Ru ions are shown, but signatures of dimerization or trimerization are not seen. (f) Liner-like chains of the Ru ions, which is a portion of the zig-zag ladder along the $b$ axis in the $Pnma$ symmetry. The three-fold periodicity of the bond length is clearly observed.} 
\label{f2}
\end{figure}

\begin{table*}[htb]
\caption{Structural parameters of RuAs in the metallic and insulating phases. The equivalent isotropic atomic displacement parameter $B_{eq}$ and the occupancy are also shown.}
\begin{center}
\begin{tabular}{cccccccc}\hline\hline
293 K & site & wyckoff & x & y & z & $B_{eq}$ (\AA$^2$) & occup. \\
\hline
&Ru & 4c & 0.00027 & 0.25 & 0.20272 & 0.59 & 1 \\
&As & 4c & 0.19557 & 0.25 & 0.56909 & 0.36 & 0.988 \\
\hline\hline
& & & & & & \\
\hline\hline
170 K & site & wyckoff & x & y & z & $B_{eq}$ (\AA$^2$) & occup. \\
\hline
&Ru1 & 4e & -0.23004(7) & 0.93147(2) & 0.26397(5) & 0.308(9) & 1 \\
&Ru2 & 4e & 0.56805(6) & 0.76233(2) & 0.56793(5) & 0.321(9) & 1 \\
&Ru3 & 4e & 0.60492(7) & 0.43577(2) & 0.60076(6) & 0.334(9) & 1 \\
&Ru4 & 4e & 0.09791(6) & 0.59599(2) & 0.59745(5) & 0.342(9) & 1 \\
&Ru5 & 4e & -0.27167(7) & 0.89775(2) & 0.73385(5) & 0.336(9) & 1 \\
&Ru6 & 4e & 0.07992(6) & 0.93010(2) & 0.58449(5) & 0.278(9) & 1 \\
&Ru7 & 4e & 0.25102(6) & 0.76444(2) & 0.25156(4) & 0.298(9) & 1 \\
&Ru8 & 4e & -0.07116(6) & 0.72837(2) & 0.43422(6) & 0.358(9) & 1 \\
&Ru9 & 4e & 0.42160(6) & 0.59933(2) & 0.91880(4) & 0.285(9) & 1 \\
&As1 & 4e & -0.28917(8) & 0.63657(3) & 0.51665(6) & 0.268(13) & 0.986(3) \\
&As2 & 4e & 0.63135(8) & 0.69241(3) & 0.81187(6) & 0.231(14) & 0.984(3) \\
&As3 & 4e & 0.05282(8) & 0.85291(3) & 0.35787(6) & 0.222(14) & 0.985(3) \\
&As4 & 4e & -0.05849(8) & 0.80788(3) & 0.65457(6) & 0.254(14) & 0.985(3) \\
&As5 & 4e & 0.39884(8) & 0.52525(3) & 0.68557(6) & 0.274(14) & 0.986(3) \\
&As6 & 4e & 0.97124(8) & 0.47511(3) & 0.65011(6) & 0.242(14) & 0.985(3) \\
&As7 & 4e & 0.37374(8) & 0.85257(3) & 0.68016(6) & 0.214(14) & 0.987(3) \\
&As8 & 4e & 0.28399(8) & 0.68482(3) & 0.48169(6) & 0.239(13) & 0.988(3) \\
&As9 & 4e & -0.28604(8) & 0.97992(3) & 0.51817(6) & 0.214(13) & 0.988(3) \\ \hline\hline
\end{tabular}
\end{center}
\end{table*}

The resultant crystal structures are shown in Fig.~4.
The structure in $Pnma$ contains the equivalent 4 Ru and 4 As atoms in the unit cell, as shown in Fig.~4(a).
In the ground state below $T_{\rm MI2}$, the system transforms to the monoclinic structure in $P2_1/c$, in which the inequivalent 9 Ru and 9 As sites exist, and totally 36 Ru and 36 As atoms are included in the unit cell.
Therefore, the unit cell in the insulating state is comparable to 9 unit cells of the original structure in $Pnma$.
The $c$ axis in $Pnma$ just corresponds to the $b$ axis in $P2_1/c$, whose length is 3 times as long as the original lattice constant.
As shown in the figure, the $ab$ plane in $Pnma$ corresponds to the $ac$ plane in $P2_1/c$.
The blue line in Fig.~4(d) indicates the unit cell of the monoclinic structure, while the red line indicates the $3 \times 3$ cell of the original unit cell.
The new three-fold periodicity can be confirmed for each direction.

Figure 4(e) shows the zig-zag chains of the Ru ions at low temperatures.
In $Pnma$, the Ru ions form the perfect zig-zag chain consisting of the equivalent sites along the $a$ axis, while it slightly deforms in $P2_1/c$.
The chain1 is composed of an alternation of 6 different Ru sites, while the chain2 includes an alternation of 3 sites.
The respective distances between the neighboring Ru ions are shown, but they are featureless and do not show a clear indication of a dimerization of the Ru ions.
Absence of a dimerization is obviously supported by an odd number of Ru sites in the unit cell, excluding a possibility of a spin-singlet formation in the insulating state.
As shown in Fig.4(f), on the other hand, the clear three-hold periodicity of the bond length can be seen along the original $b$ axis, where the bonding built a liner chain with equal intervals of 3.328 \AA \ in the metallic phase.
In a linear chain, for example, the Ru1-Ru6 and Ru3-Ru6 bondings shrink to $\sim3.1$ \AA, while the Ru1-Ru3 bonding significantly extends to $\sim3.66$ \AA. This tendency is commonly seen for all the bondings along the original $b$ axis. This key feature in the deformed structure will be discussed in the final part, together with results of a band calculation.

\subsection{Nuclear quadrupole/magnetic resonance}

\begin{figure}[htb]
\begin{center}
\includegraphics[width=\linewidth]{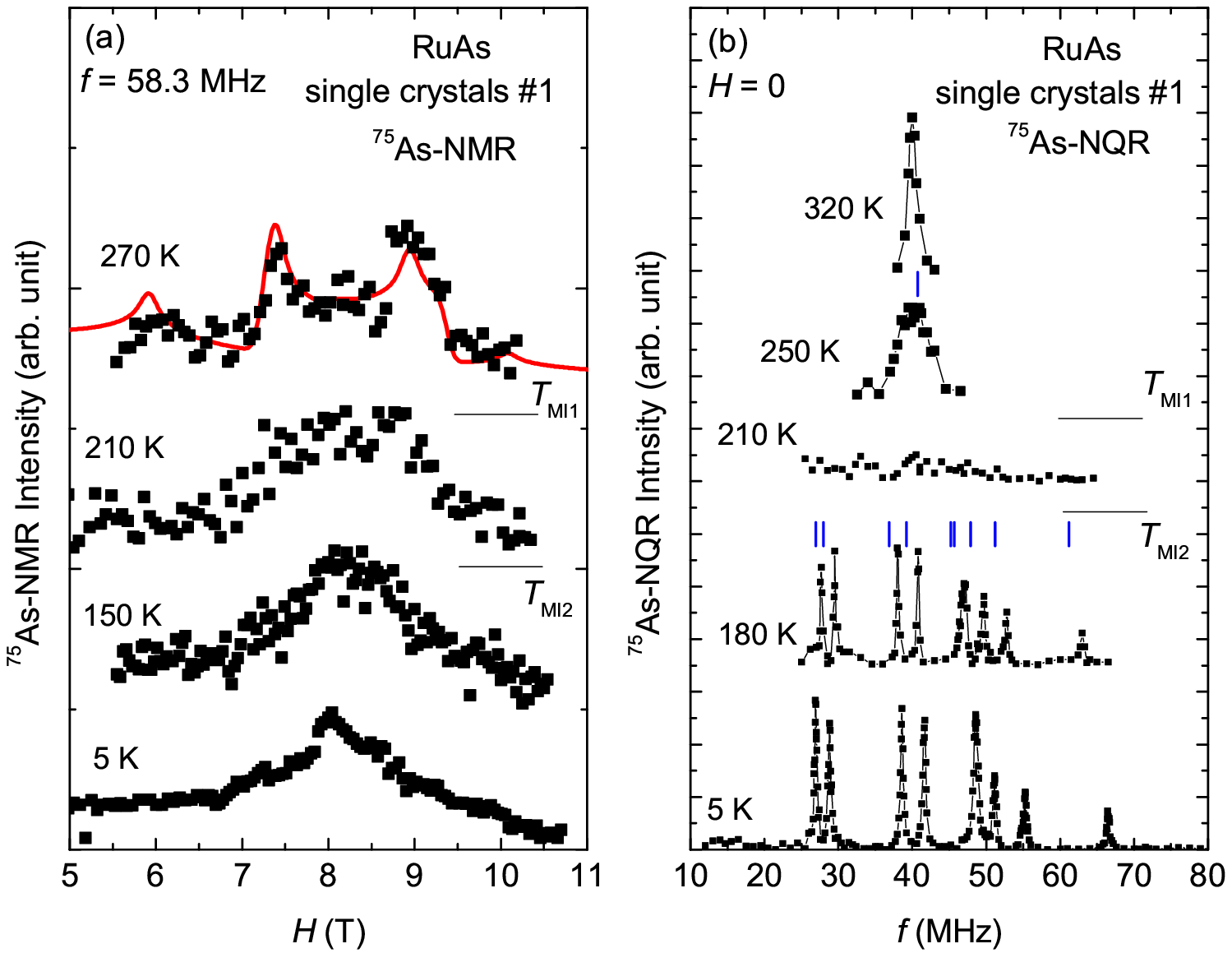}
\end{center}
\caption{(color online) (a) Field-swept NMR spectrum for RuAs measured at around 8 T. (b) NQR spectrum for RuAs measured at zero field. In the metallic state above $T_{\rm MI1}$, the NMR spectrum obtained using many pieces of single crystals shows a typical powder pattern by assuming the unique As site, as shown by the red curve. This is consistent with the NQR spectrum observed at around 40.0 MHz. Below $T_{\rm MI1}$, the broadened NMR spectrum and the suppression of the NQR intensity indicate the strong distribution of $\nu_Q$ and $\eta$. In the ground state below $T_{\rm MI2}$, the several peaks are observed at the NQR measurement, indicative of the formation of the superlattice. The blue lines indicate the resonance frequencies calculated using the structural parameters at 293 K and 170 K.}
\label{f2}
\end{figure}

We performed NMR/NQR measurements to investigate the microscopic state of RuAs.
Figures 5(a) and (b) show the NMR spectrum measured at around 8 T and the NQR spectrum measured at zero field, respectively.
We used many pieces of small single crystals for both measurements; therefore, the experimental condition is similar to a measurement using polycrystalline samples.
Here, the NMR/NQR data using the crystal \#1 are shown, but the sample dependence for the spectral shape was negligible. 
In Fig.~5(a), we observed the broad spectra near the position of zero Knight shift ($H\sim8$ T) for each temperature.
Here, we could not estimate Knight shift owing to the spectral broadening by the quadrupole interaction.
At 270 K in the metallic state, the spectrum can be reproduced by a simulation shown by a red curve as a typical powder pattern affected by a quadrupole interaction.
We obtained the quadrupole frequency $\nu_Q=39.8$ MHz and the asymmetry parameter $\eta=0.2$, which give the resonance frequency of $\nu_{res}=40.1$ MHz at zero field.
As shown in Fig.~5(b), we observed the NQR signal forming a single peak at around 40.0 MHz, which corresponds to $\pm1/2 \leftrightarrow \pm3/2$ transition.
All the As sites are equivalent in the crystal structure represented by the space group of $Pnma$, which is consistent with these observations.
In the NQR spectrum, the spectral width is quite narrow at a high temperature of 320 K, but it is significantly broadened just at $T_{\rm MI1}=250$ K.
In the intermediate temperature range between $T_{\rm MI2}$ and $T_{\rm MI1}$, the NQR spectrum is strongly broadened and almost disappears.
The contrastive observation of the NMR signal in the same phase, for which the Zeeman interaction is predominant, excludes a possibility that the fast relaxation weakens the NQR intensity.
Therefore, the electric field gradient (EFG) parameters, $\nu_Q$ and $\eta$ are suggested to be distributed strongly in this phase.
This is in sharp contrast to the ground state below $T_{\rm MI2}=190$ K, where the sharp signals for the NQR measurement are suddenly recovered, and we observed 8 resonance lines.
From this result and the structural analysis, it is clear that the ground state of RuAs is commensurate. 
On the other hand, the broad and weak NQR intensity in the intermediate phase implies that the charge distribution is incommensurate.

\begin{table}[htb]
\caption{Calculated and experimental EFG parameters of RuAs in the metallic and insulating phases.}
\begin{center}
\begin{tabular}{cccccc}\hline\hline
\multicolumn{6}{c}{metallic phase} \\ 
\hline
& \multicolumn{3}{c}{293 K} & 290 K & \\
site & $\nu_Q^{cal}$ & $\eta^{cal}$ & $\nu_{res}^{cal}$ & $\nu_{res}^{exp}$ & error \\
& (MHz) & & (MHz) & (MHz) & (\%) \\
\hline
As & 36.7 & 0.833 & 40.7 & 40.0 & -1.8 \\
\hline\hline
& & & & \\
\hline\hline
\multicolumn{6}{c}{insulating phase} \\
\hline
& \multicolumn{3}{c}{170 K} & 180 K & \\
site & $\nu_Q^{cal}$ & $\eta^{cal}$ & $\nu_{res}^{cal}$ & $\nu_{res}^{exp}$ & error \\
& (MHz) & & (MHz) & (MHz) & (\%) \\
\hline
As6 & 25.68 & 0.551 & 26.95 & 27.6 & +2.4 \\
As2 & 25.35 & 0.810 & 27.99 & 29.5 & +5.1 \\
As7 & 35.39 & 0.508 & 36.89 & 38.0 & +2.9 \\
As9 & 37.65 & 0.508 & 39.24 & 40.9 & +4.1 \\
As1 & 40.70 & 0.837 & 45.20 & 47.0 & +3.8 \\
As8 & 43.47 & 0.562 & 45.70 & 47.0 & +2.8 \\
As3 & 47.13 & 0.317 & 47.91 & 49.7 & +3.6 \\
As4 & 48.76 & 0.554 & 51.19 & 52.7 & +2.9 \\
As5 & 54.71 & 0.869 & 61.22 & 62.9 & +2.7 \\ \hline\hline
\end{tabular}
\end{center}
\end{table}

Next, we checked the resonance frequencies at the inequivalent As sites through a first principle calculation of the EFG using the structural parameters.
The calculation method of EFG has been explained elsewhere,\cite{Blaha} and it has been used successfully for many systems.\cite{Blaha2,Schwarz,Rusz,Harima_EFG}
Here, the nuclear quadrupole moment for the As nucleus, $Q=314 \times 10^{-31}$ m$^2$ is utilized to convert EFG to $\nu_Q$.\cite{Q}
Table III shows $\nu_Q^{cal}$, $\eta^{cal}$, and the resultant resonance frequency $\nu_{res}^{cal}$ estimated from the calculation, and $\nu_{res}^{exp}$ observed experimentally.
In the metallic state, the calculated $\nu_Q^{cal}$ is close to the experimental $\nu_{Q}^{exp}=39.8$ MHz estimated from the NMR spectrum, while $\eta^{cal}$ is deviated from the experimental $\eta=0.2$.
Since $\nu_{res}$ is not so sensitive to $\eta$, $\nu_{res}$ has a good consistency between the calculation and the experiment within the small error, which is defined by ($\nu_{res}^{exp} - \nu_{res}^{cal}$) / $\nu_{res}^{exp}$.
In the insulating state, $\nu_Q^{cal}$, $\eta^{cal}$, and $\nu_{res}^{cal}$ for all 9 As sites are calculated and shown in ascending order of $\nu_{res}^{cal}$ in the table.
The experimental $\nu_{res}^{exp}$ is also listed in ascending order. 
In the NQR spectrum shown in Fig.~5(b), we observed 8 inequivalent signals; therefore, 1 site is missing.
The intensity analysis in the wide frequency range is difficult to keep an accuracy, but the signal just below 50 MHz has the largest intensity (area) and the broadest width among the 8 signals, indicating that 2 peaks are accidentally overlapped in the signal.
This interpretation is strongly supported by the calculation, in which the As1 and As8 sites have the almost same resonance frequencies, which are slightly smaller than the subjected resonance frequency.
In this context, the $\nu_{res}^{exp}$ for all the As sites show the almost regular error of about $+2.4\sim5.1$ \% for the calculation. 
The resultant $\nu_{res}^{cal}$ are shown by the blue lines in Fig.~5(b) for each phase.
The present calculation reproduces excellently the EFG at the As sites, ensuring the high validity of the obtained structural parameters as well as the reliability of the calculation.
Simultaneously, this result reveals that the ground state of RuAs is nonmagnetic, because the spectral splitting by the internal field can be ignored.

\begin{figure}[ht]
\centering
\includegraphics[width=\linewidth]{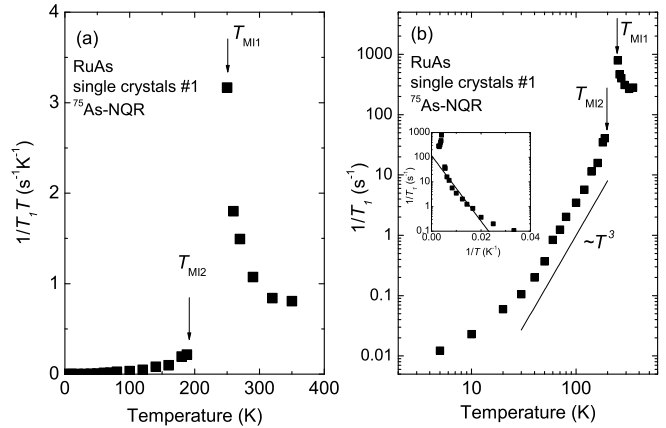}
\caption[]{(color online) Temperature dependence of (a) $1/T_1T$ and (b) $1/T_1$ for RuAs. RuAs undergoes the insulating state accompanied by a strong development of some fluctuations. The $T^3$ behavior below $T_{\rm MI2}$ indicates that the energy gap of RuAs is anisotropic.
}
\end{figure}

Figure 6(a) shows the temperature dependences of $1/T_1T$ for the single crystal RuAs.
$T_1$ was measured at the peak of $\sim49$ MHz in the insulating state, and at the single peak at $\sim40$ MHz in the metallic state.
Unfortunately, we could not measure $T_1$ in the intermediate region between $T_{\rm MI1}$ and $T_{\rm MI2}$ owing to the weak intensity.
The $1/T_1T$ shows a strong divergence toward $T_{\rm MI1}$ and obvious suppression below $T_{\rm MI2}$.
The divergence shows that a symmetry lowering by the metal-insulator transition is accompanied by strong critical fluctuations.
Since the As nuclear spin is $I=3/2$, both magnetic and electric relaxations are possible for the origin of the divergence of $1/T_1$.
The absence of Curie-Weiss like behavior in magnetic susceptibility suggests that the magnetic correlations of $\vec{q}=0$ is weak in this system, but the concerned wave vector is probably finite, because the superlattice of $3 \times 3 \times 3$ is formed at the low temperatures.
$1/T_1T$ corresponds to a $\vec{q}$-summed dynamical susceptibility.
Therefore; the presence of magnetic correlations cannot be excluded experimentally at present. 
Figure 6(b) displays the temperature dependence of $1/T_1$ for RuAs.
$1/T_1$ in RuAs follows $T^3$ behavior below $T_{\rm MI2}$ in contrast to a conventional exponential behavior.
One possibility is that the relaxation at low temperatures is extrinsically dominated by a small amount of magnetic impurities. We cannot exclude this possibility because the Curie tail appears in the susceptibility at low temperatures. 
Another possibility is intrinsically that the density of states (DOS) near the Fermi energy, $E_F$ has a linear energy dependence, that is, $D(E') \sim |E'|$, where $E'=E-E_F$, unlike a simple full gap. 
The thermal excitation in this energy dependence gives $T^3$ dependence in $1/T_1$. 
If this is the case, the band structure near $E_F$ is expected to be like a semimetal. 
We cannot judge which is correct, but an estimation of the energy-gap was tried for a comparison by the same manner as that attempted for RuP.\cite{Li}
Using $1/T_1 \sim \exp(-E_g/k_BT)$ for the experimental data between $T_{\rm MI2}$ and 50 K, as shown in the inset of Fig.~6(b), we obtained $E_g =308$ K $\pm 30$ K, which is comparable to 340 K estimated from the resistivity in the same temperature range.
The energy gap of RuAs is several times smaller than $1250$ K for RuP,\cite{Li} which is qualitatively consistent with a difference in the resistivity between two compounds.\cite{Hirai}

\subsection{Electronic structure calculation}

\begin{figure}[b]
\centering
\includegraphics[width=0.85\linewidth]{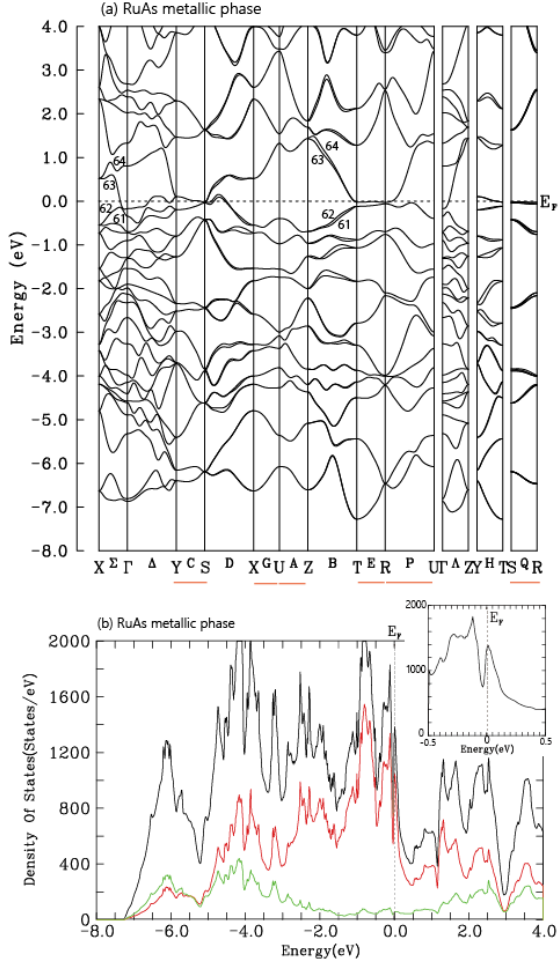}
\caption[]{(color online) The calculated energy dispersion and the DOS of RuAs in the metallic phase in the orthorhombic $Pnma$ space group. The four-fold degeneration is realized along axes shown by red lines in the Brillouin zone boundary even though the spin-orbit coupling is considered, because of the protection by the non-symmorphic symmetry. Two hole bands and two electron bands denoted by $61-64$ cross the Fermi level. The flat bands near $E_F$ are seen along the degenerated $Y-S$, $T-R$, and $S-R$ axes, producing the large DOS at $E_F$. 
In the lower panel, the black curve indicates the total DOS integrated in the whole $k$ space. The red and green curves indicate the partial DOS originating from the Ru-$4d$ and As-$4p$ orbital, respectively.
}
\end{figure}

\begin{figure}[h]
\centering
\includegraphics[width=0.85\linewidth]{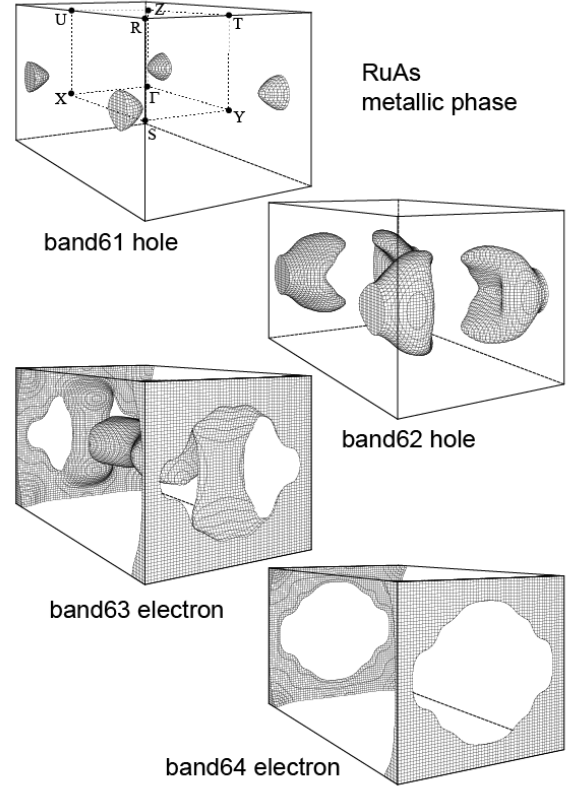}
\caption[]{The Fermi surfaces of RuAs in the metallic phase in the $Pnma$ space group. The 63 and 64 electron sheets are almost degenerated and low dimensional.
}
\end{figure}

Figures 7 (a) and (b) show the calculated energy dispersion and the DOS for the metallic phase in the $Pnma$ space group, and the Fermi surfaces are drawn in Fig.~8.
In the energy dispersion, the flat bands near $E_F$ are seen along the $T-R$, $S-R$ and partially $Y-S$ axes, and they construct the peak structure in the DOS near $E_F$.
This feature is consistent with the previous calculation.\cite{Goto} 
The spin-orbit coupling is taken into account in the present calculation, and the DOS is slightly modified and the peak feature near $E_F$ became more remarkable. 
The band splitting by the spin-orbit coupling can be seen along the specific axes, such as $S-X$, $Z-T$, and $Y-T$, where only two-fold spin degeneracy remains.
However, the four-fold degeneracy is completely maintained, for example, along the $Y-S$, $T-R$, and $S-R$ axes, which are on the Brillouin zone boundary, as indicated by red lines in Fig.~7(a), because it is protected by the non-symmorphicity of the $Pnma$ space group.\cite{Niu}
This feature will be a common point to contribute an electronic state of systems in the non-symmorphic $Pnma$ space group, for instance, it can be a key ingredient to interpret a magnetoresistance of the isostructural CrAs.\cite{Niu}
This protection yields the almost degenerate Fermi surfaces at the Brillouin zone boundary in the metallic phase of RuAs.
The plate-like Fermi surfaces with a hollow in the middle are seen in 63 and 64 electron bands, and they partially degenerate because of the protected degeneracy along the $Y-S$, $T-R$, and $S-R$ axes. 
Interestingly, the flat bands near $E_F$ exist along the axes with the four-fold degeneracy protected by the crystal symmetry.
This certainly produces the peak structure in the DOS and makes this phase unstable at low temperatures; therefore, the lifting of this degeneracy by a symmetry lowering from the non-symmorphic $Pnma$ is a key ingredient of the metal-insulator transition.

\begin{figure}[tb]
\centering
\includegraphics[width=0.85\linewidth]{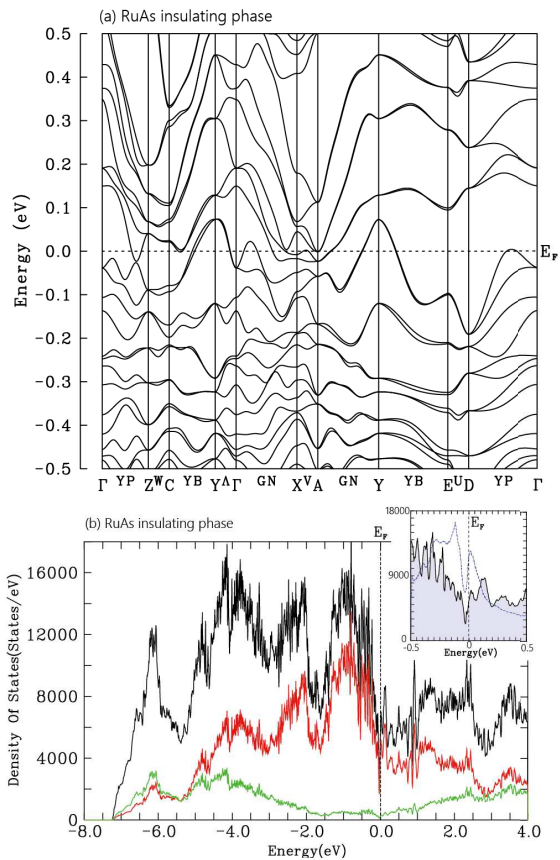}
\caption[]{(color online) The calculated energy dispersion in the vicinity of $E_F$ and the DOS of RuAs in the insulating phase in the monoclinic $P2_1/c$ space group. A gap-like feature is likely to appear in the DOS near $E_F$. In the lower panel, the black curve indicates the total DOS integrated in the whole $k$ space. The red and green curves indicate the partial DOS originating from the Ru-$4d$ and As-$4p$ orbital, respectively. In the inset, the solid line with the hatching region indicates the total DOS for the insulating state, while the dotted line shows the metallic state. Here, the DOS of the metallic phase is multiplied by 9 to adjust the difference in the size of the unit cell.
}
\end{figure}

We test a band calculation for the ground state in the $P2_1/c$ space group, and the energy dispersion in the vicinity of $E_F$ and the DOS are shown in Fig.~9.
Here, the Brillouin zone is that of the monoclinic structure.
We can confirm a disappearance of the degenerate flat bands, but the present calculation did not produce a clear energy gap, which is relatively small as estimated to be $E_g=340$ K $\sim$ 0.03 eV from the resistivity.
The result shows a semimetallic band dispersion as an important feature; two electron bands and two hole bands cross $E_F$, and a switching of the electron bands and the hole bands does not occur in the momentum space.
As a result, a valley of DOS near $E_F$ approximates a gap-like feature, as shown in the inset of Fig.~9(b).
The DOS at $E_F$ of $\sim5600$ states/eV is significantly reduced from $\sim12000$ states/eV for the 9 unit cells in the metallic phase.
However, the absence of the clear gap in the calculation indicates a presence of some lacking factor to reproduce the electronic state of the ground state completely.
Generally, the LDA calculation tends to underestimate the energy gaps of semiconductors or insulators.\cite{Ashcroft,Zheng}
This is interpreted as inaccuracy in the approximation itself; therefore, it may be difficult to reproduce the small energy gap of $\sim0.03$ eV in the LDA framework.
In other words, inadequate consideration of Coulomb interaction causes the underestimation of the gap.
The energy gap in principle may open by increasing Coulomb interaction numerically in the semimetallic electronic state, although the energy shift of approximately $0.1$ eV between the electron and hole bands is required in this case.
In any case, the development of the gap-like feature accompanied by the semimetallic band structure suggests that the lifting of the degeneracy through the formation of the superlattice promotes to produce the insulating state.
In this sense, the transition is caused by a band Jahn-Teller effect.

\begin{table}[htb]
\caption{Calculated $d$-electron numbers at each Ru site in the insulating phase.}
\begin{center}
\begin{tabular}{ccc}
\hline
\multicolumn{3}{c}{insulating phase} \\
\hline
site & $d$-electron & a difference from \\
& number & the average (\%) \\
\hline
Ru1 & 5.301 & +0.038 \\
Ru2 & 5.305 & +0.113 \\
Ru3 & 5.303 & +0.075 \\
Ru4 & 5.310 & +0.208 \\
Ru5 & 5.301 & +0.038 \\
Ru6 & 5.291 & -0.151 \\
Ru7 & 5.290 & -0.170 \\
Ru8 & 5.304 & +0.094 \\
Ru9 & 5.286 & -0.245 \\ \hline
\end{tabular}
\end{center}
\end{table}

\begin{figure}[htb]
\centering
\includegraphics[width=0.95\linewidth]{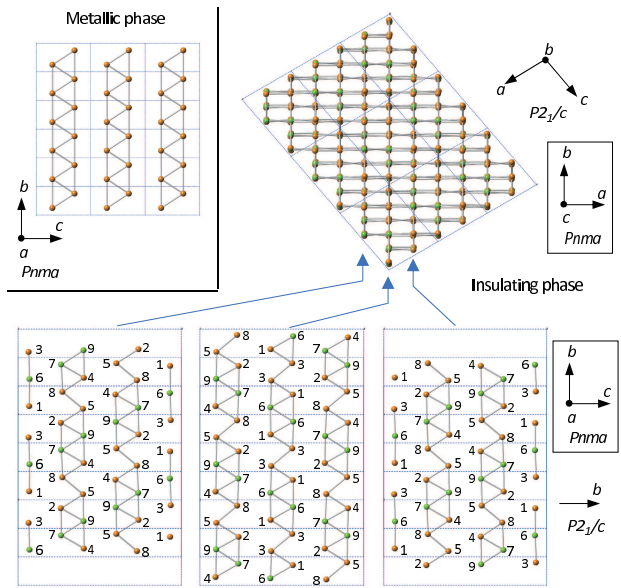}
\caption[]{(color online) The arrangement of the Ru ions in the metallic state (upper left) and the insulating state (upper right and lower). In the metallic phase, the Ru ions form the zig-zag ladder along the $b$ direction. In a lower structure, the number indicates each Ru site. The Ru6, Ru7 and Ru9 sites, where the calculated electron numbers are fewer, are shown by green spheres. In the $bc$ plane of the original $Pnma$ structure, a linear alignment of those sites is realized. In the lower structure, the shorter Ru-Ru bonding of less than 3.21 \AA \ are drawn. 
}
\end{figure}

Another important point is why the superlattice of $3 \times 3 \times 3$ is necessary at low temperatures, because the superlattice formation is not essential to lift the degeneracy protected by non-symmorphic symmetry.
RuAs itself is a compensated metal in the $Pnma$ space group, and it can be insulating with maintaining the size of the unit cell.
For example, the simple dimerization along the zig-zag chain (along the $a$ axis in $Pnma$), which does not change the size of the unit cell, can lift this degeneracy, but this is not realized.
To check a relationship between the structural modulation and the electronic state, we evaluated the $d$-electron number of each Ru site in the insulating state, which was obtained in the band calculation, and they are shown in Table IV. 
The average number of $d$ electrons at all the Ru sites is 5.299, which corresponds to Ru$^{2.701+}$.
The differences from the average number at each Ru site are also listed in the table.
The $d$ electron numbers at the Ru6, Ru7, and Ru9 sites are obviously fewer than those of other sites.
An arrangement of these Ru ions in the crystal structure is shown in Fig.~10, where the Ru6, Ru7, and Ru9 sites are drawn by green spheres and the As sites are eliminated.
An upper right structure is the $ac$ plane in the $P2_1/c$ space group, which is equivalent to that in Fig.~4(d) and a lower figure shows each layer normal to the $ac$ plane.
These layers are the $bc$ plane in the original $Pnma$ space group, and it consists of the zig-zag ladders.
As shown in the upper left structure, the zig-zag ladder was constructed by the equivalent Ru sites in the metallic phase, and it deforms in the insulating state.
The shorter Ru-Ru bondings, whose length is less than 3.21 \AA, are shown in the lower figure.
As already shown in Fig.~4, some bondings obviously extend, making a clear three-hold periodicity along the original $b$ axis.
The Ru6, Ru7, and Ru9 sites possess 4 shorter bondings in the layer, while other sites have 3 shorter bondings.
This difference is likely to induce the distribution of the $d$ electron numbers.
Interestingly, the Ru6, Ru7, and Ru9 sites align in a straight line in this layer, indicative of a formation of a stripe-type charge modulation.
Reflecting the original zig-zag chain along the $a$ axis in the $Pnma$ space group, which is perpendicular to these layers, the direction of the stripe changes alternatively layer by layer.
This apparent coupling of the structural and electronic modulation suggests a formation of commensurate CDW in the ground state of RuAs.
It is conjectured that the $3\times3\times3$ superlattice is required to produce this structural and electronic modulation.
Numerical studies are required to elucidate why this modulation is the most stable at low temperatures. In this context, a contribution of the nesting of the Fermi surfaces is also a matter of concern. The nesting property with the corresponding wave vector will be checked by a numerical analysis in the future.

\section{Summary}

In summary, the several experiments have been performed for the single crystals of RuAs obtained successfully using Bi-flux method. 
It shows clear successive metal-insulator transitions at $T_{\rm MI1}\sim255$ K and $T_{\rm MI2}\sim195$ K alike the polycrystalline sample.
This enables us to escape from an awkward problem on the sharp difference between the poly- and single- crystals of RuP.
The X-ray structural analysis and the NQR spectrum for RuAs demonstrate the nonmagnetic superlattice of $3\times3\times3$ is formed in the ground state, while the detailed structure and electronic state in the intermediate phase are unclear yet.
Although the decided superlattice structure of the ground state is complicated, the validity is guaranteed strongly through the excellent agreement of the quadrupole frequencies between the experiment and the calculation.
The revealed crystal structure does not show a clear indication of dimerization of the Ru ions, and an odd number of Ru sites in the unit cell excludes a possibility of a spin-singlet formation.
The structural modulation and its clear coupling with the calculated $d$ electron number suggests a formation of a stripe-type CDW, where the direction of the stripe alternates layer by layer. 
A direct observation of this electronic modulation is an important issue, and a verification of the orbital state at each Ru site is also an interesting subject.
RuAs is not a simple low-dimensional material but possesses the Fermi surface instability caused by a non-symmorphicity of the $Pnma$ space group, resulting in the transition to the insulating state with the characteristic structural and electronic modulation.

\section*{Acknowledgement}

The authors thank Y. Kuramoto, S. Kimura, D. Hirai, H. Ikeda, and A. Koda for helpful discussions.
This work has been supported in part by Grants-in-Aid for Scientific Research (Nos. 15H03689, 15H05745, and 26400359) from the Ministry of Education, Culture, Sports, Science and Technology (MEXT) of Japan.

\end{document}